# Large-Area and Transferred High-Quality Three-Dimensional Topological Insulator $Bi_{2-x}Sb_xTe_{3-y}Se_y$ Ultrathin Film by Catalyst-Free Physical Vapor Deposition

*Ngoc Han Tu[†], Yoichi Tanabe[†\*], Yosuke Satake[†], Khuong Kim Huynh[‡], Le Huu Phuoc[†], Stephane Yu Matsushita[†], and Katsumi Tanigaki[†, ‡\*\*]*

[†]Department of Physics, Graduate School of Science, Tohoku University, Sendai, 980-8578, Japan.
[‡]WPI Advanced Institute for Materials Research, Tohoku University, Sendai 980-8577, Japan.

Email: ytanabe@m.tohoku.ac.jp, tanigaki@m.tohoku.ac.jp
TEL : +81-22-217-6173, FAX : +81-22-795-6470

ABSTRACT

Uniform and large–area synthesis of bulk insulating ultrathin films is an important subject toward applications of a surface of three dimensional topological insulators (3D-TIs) in various electronic devices. Here we report epitaxial growth of bulk insulating three-dimensional topological insulator (3D-TI) $Bi_{2-x}Sb_xTe_{3-y}Se_y$ (BSTS) ultrathin films, ranging from a few quintuple to several hundreds of layers, on mica in a large–area (1 $cm^2$) via catalyst-free physical vapor deposition. These films can nondestructively be exfoliated using deionized water and transferred to various kinds of substrates as desired. The transferred BSTS thin films show good ambipolar characteristics as well as well-defined quantum oscillations arising from the topological surface states. Carrier mobility of 2500 – 5100 $cm^2$/Vs is comparable to the high quality bulk BSTS single crystal. Moreover, tunable electronic states from the massless to the massive Dirac fermion were observed with a decrease in the film thickness. Both the feasible large area synthesis and the reliable film transfer process can promise that BSTS ultrathin films will pave a route to many applications of 3D-TIs.

A new concept of topological insulators proposed for 2D graphene with time-reversal $Z_2$ invariance [1] and its clear experimental evidence recently given for cadmium telluride with superlattice structure [2] as well as other three dimensional topological insulators (3D-TIs) have been giving a strong impact in materials science [3-5]. A spin momentum locked Dirac cone surface of 3D-TIs is predicted as a source of an ultralow dissipative spin-charge current where the electron's spin orientation is relatively locked to its propagation direction. Because this current induced spin polarization is efficient to rewrite a magnetic memory under the least possible current and energy consumption, the 3D-TI is promising for the future spintronics and therefore various theoretical and experimental researches were conducted to manipulate its intrinsic physical properties[6-25].

For practical usages of 3D-TIs, however, high quality 3D-TI surface with a tunable Fermi level under bulk insulating electronic states as well as large surface area on various targeted substrates are required. For this purpose, mechanical exfoliation[9, 11, 26-28], solvothermal synthesis[29], chemical vapor deposition (CVD) [30, 31], and physical vapor deposition (PVD)[32-34] have been studied to date, but uniform 3D-TI ultrathin films with bulk insulating electronic background states are only available in the limited size (usually less than 100 μm). Although molecular beam epitaxy (MBE) [13, 35, 36] has frequently been employed for a large-area synthesis of bulk insulating 3D-TI thin films in the fundamental research, both time-consumed processes and expensive production cost are the serious bottleneck for practical applications. Moreover, a controllable growth of some important bulk insulating 3D-TI compounds such as $Bi_{2-x}Sb_xTe_{3-y}Se_y$ (BSTS) was not achieved by the MBE promising for the future application studies based on the splendid bulk insulating properties[37, 38] and the robust quantum Hall state[12] in the bulk single crystal. Consequently, uniform and large–area synthesis of bulk insulating 3D-TI ultrathin films is presently the important priority research.

In this report, we demonstrated that high quality bulk insulating 3D-TIs can epitaxially be grown in centimeter-size, via catalyst-free physical vapor deposition (CF-PVD) for BSTS [37, 38] on a mica substrate by controlling nucleation and growth accurately. Importantly, the grown ultrathin films can be exfoliated from mica using deionized water and nondestructively transferred to any other substrates, such as silicon, without losing the original high quality Dirac surface states with high carrier mobility. An ambipolar electronic nature of Dirac fermions is demonstrated in a field effect transistor by applying back gate voltage on a $SiO_2$/Si substrate. Magnetoresistance showing Shubnikov–de Haas (SdH) oscillations under high magnetic fields exhibits a half-integer Berry phase shift in the Landau level fan diagram plot and clearly confirms the nontrivial topological surface. Carrier mobilities of 2500 $cm^2V^{-1}S^{-1}$ and 5100 $cm^2V^{-1}S^{-1}$ for the top and the bottom surfaces are evaluated from the SdH oscillations and these are the highest among 3D-TI films so far transferred on Si substrates [39]. Moreover, when the thickness was reduced to the ultrathin region, an on/off switching ratio in the field effect transistor was dramatically enhanced to ~ $10^4$, being consistent with the emergence of the massive Dirac fermion. The present growth and film transfer method of high quality 3D-TI ultrathin film will be an important mile stone towards practical usages of 3D-TIs in novel electronic/spintronic devices.

Considering that BSTS (**Fig. 1(a)**) can be one of the best candidates for topological insulators for various applications thanks to the splendid bulk insulating properties, we challenged to grow a uniform ultrathin film with high quality in a large–area. **(See Supporting Information Section1 and Fig. S1)** Because of a small lattice mismatch, fluorophlogopite mica ($KMg_3(AlSi_3O_{10})F_2$) is reported to be a good substrate for van der Waals epitaxy of $Bi_2Se_3$ and $Bi_2Te_3$ [33] and can be applied for single crystalline nanoplate growth of BSTS with a catalyst-free condition [40]. We equipped a double quartz tube system (**Fig. 1(b)**) in order to have uniform and large–area synthesis of high crystalline ultrathin film with a variable thickness on a substrate by making accurate control in growth condition as described later. It is important to grow BSTS TI thin films on a different substrate because other substrates, such as SiO$_2$, are required for device construction. However, high quality epitaxial growth is generally difficult when the lattice matching condition is not fulfilled. Alternatively, we importantly found that the grown BSTS ultrathin films can be exfoliated and transferred to other substrates without any damage using a deionized water as displayed in **Fig.1(c)-(f)**.

In the initial stage of growth, hexagonal or triangular - shaped BSTS nanoplates (NPs) were uniformly nucleated on mica with similar orientation (**Fig. 2**, also **See Supporting Information Fig. S2**). As the total growth time increased, BSTS NPs enlarged in the lateral direction to the substrate and larger NPs were grown. A one-end-closed quartz tube of inner-diameter of 0.4 - 1.5 cm placed inside a larger quartz tube of 3 cm inner diameter in a furnace (**Fig. 1(b)**) was very effective to control the crystal growth. A source material was placed at the upstream position (ca. 500 °C) of the smaller tube located at the closed-end of tube while a mica substrate was set at the downstream position of the open-end of tube (ca. 360 °C). The tube was evacuated down to 0.1 Pa and refilled with Ar gas to remove the residual oxygen repeatedly for several times followed by crystal growth under continuous evacuation of 0.1 Pa. Morphology and thickness of BSTS films were strongly influenced by the growth time and the quality of the source materials. Uniform BSTS films with a thickness of 8 nm were grown in a 1 cm$^2$ area after 10 min growth as shown in the photograph image (**Fig.1(c)**).

Interestingly, two different thin film growth modes were found depending on the inner tube diameter of $d$ between 1 and 1.5 cm, as shown in **Fig. 2**. In $d$ = 1 cm as shown in **Fig. 2 (a)** (**See Supporting Information Section 2 and Fig. S2**), hexagonal or triangular - shaped NPs with ~ 10 μm in size and thickness of 12 nm were grown for 15 min, ending up with larger NPs via the layer–plus–island Stranski-Krasrtanov growth mode [41]. Upon prolonging the growth time, another additional layer was gradually grown on the first layer. On the other hand, in $d = 1.5$ cm as shown in **Fig. 2 (b) and (c)** (**See Supporting Information Section 2 and Fig. S3**), a BSTS film composed of well oriented BSTS NPs with thinner thickness of 3 nm was obtained after 5 min growth. By increasing the growth time, these two – dimensional NPs with several μm in size stacked on the precedently grown plates, ending up with a large single crystal NP in the layer–by–layer van der Waals epitaxy growth mode [42]. The latter growth mode is desirable for the purpose of TI applications to electronic devices.

Taking various applications of 3D-TI thin films into account, the grown ultrathin films on mica in the layer–by–layer mode were exfoliated and transferred on other substrates. We tested various exfoliation methods of BSTS thin films grown on mica and found a newly effective method that can preserve the original crystal structure and the mobility of BSTS as demonstrated in **Fig.1 (d, f).**

As–grown films were dipped into deionized water in several to several tens of seconds **(Fig. 1(d))**. For the thicker film ($t > 15$ nm), as-grown BSTS thin films were precedently dipped into a 2.38 % tetra-methyl ammonium hydroxide (NMD–W 2.38%) aqueous solution in several seconds **(Fig. 1(f))** until the outer edge of BSTS film began to be peeled off from the substrate **(Fig. 1(f))**. NMD–W was chemically inert to BSTS. A completely peeled BSTS thin film with a $t$ range from 3 nm to 30 nm floats on a water surface as a free–standing film (**Fig. 1(d-f)**, also **See Supporting Information Fig. S4**) after dipping, was transferred on a $SiO_2$ substrate **(Fig. 3(a-d))**. For $t > 15$ nm, NMD–W was washed out by deionized water, acetone, isopropanol. Finally, the transferred films were subjected to one hour annealing at 100°C in a vacuum oven to remove residual water. In the earlier report, the as-grown 3D-TI thin film was covered by the polymethyl methacrylate (PMMA) then the buffered oxide etching (BOE) or potassium hydroxide (KOH) solution[39] or KOH solution and deionized water[43] was employed to peel off the 3D-TI film from a sapphire substrate. However, the carrier mobility was suppressed to be $900 - 1800$ $cm^2V^{-1}S^{-1}$ [39] and the PMMA is hard to remove perfectly[44]. Therefore, the present transferred free–standing BSTS films preserving the original film shape without such protection is a useful method to fabricate various electronic devices using 3D-TI ultrathin films.

BSTS thin film epitaxially grown on mica was transferred to a SiO$_2$/Si substrate after exfoliation and characterized by energy dispersive X–ray (EDX) spectroscopy, X-ray photo emission spectroscopy (XPS), X–Ray diffraction (XRD) and Raman spectroscopy. For EDX measurements, the as-grown film on mica was first transferred on a conducting carbon tape or a SiO$_2$/Si (300 nm) substrate, because selenium (Se) ratio was usually underestimated compared to other elements in the case of mica substrate. EDX analyses revealed that BSTS films with the uniform chemical composition of Bi:Sb:Te:Se of 1.53:0.47:1.71:1.29 in atomic ratio were grown within the resolution of EDX **(See Supporting Information Fig.S6).** XPS showed that the surface oxidization for the as-grown film tends to be comparable to previous reports in Bi$_2$Se$_3$ nanoribon[45] and no additional surface oxidization in the transferred film (**See Supporting Information Section 3 and Fig. S5**). After transferring as-grown BSTS films with 30 and 18 nm in thickness onto a glass substrate, their structures were reexamined by XRD. Clear (00ℓ) (ℓ=3n, where n in integer) diffraction peaks of BSTS (**Fig. 3(e)**) were observed. The Raman spectra showed the same peak positions before and after the film transfer processes with keeping the same five peaks at 109, 125, 136, 156, 163 cm$^{-1}$ shown in **Fig. 3(f)**, being consistent with the previous results [40] (**Also See Supporting Information Fig. S6**). Thin films from 3 to 30 nm in thickness were evaluated in the same fashion by Raman spectra as shown in **Fig. 3(g)**. It is apparent that BSTS thin films can nondestructively be exfoliated from mica and transferred to other substrates in keeping the as-grown compositions.

In order to confirm the electronic states of the topological surface of the BSTS thin films, electrical transport was measured using various thickness films. Temperature dependence of the longitudinal resistance $R_{xx}$ systematically changed with a decrease in a film thickness, being consistent with previous reports [25,46] (**See Supporting Information Section 4 and Fig. S7**). After the film transfer from mica to a Si substrate with an oxide layer of 300 nm, we fabricated a bottom–gated BSTS thin film transistor with six Au/Cr (50nm/5nm) electrodes by thermal evaporation using a metal mask. We selected a film with 3 nm, 10 nm, 18 nm in thickness to demonstrate accurate tunability of its chemical potential by controlling back gate voltage ($V_G$). As shown in **Fig.4 (a),** ambipolar carrier conduction was clearly observed by controlling $V_G$ for the 18 nm thick film. As $V_G$ becomes large to the negative direction, $R_{xx}$ (black line) increased until reaching a maximum at $V_G = -130$ V and then decreased. The same situation can also be viewed for the Hall coefficient (red line) in the same figure, where the sign changed from minus (electron conduction) to plus (hole conduction). For the 10 nm thick film as shown in **Fig. 4(b)**, an on/off ratio in $R_{xx}$-$V_G$ curve tended to be enhanced, being consistent with a decrease in a contribution of bulk electronic states. A clear two peak structure at around - 100 V and -130 V indicated the non-equivalent Fermi level shift for the top and bottom surface states in the bottom–gated 3D-TI thin film transistor [12,25]. The $V_G$ dependence of $R_{xx}$ was a little bit asymmetric in shape before and after the sign change, and such $R_{xx}$-$V_G$ characteristics resemble the previous observations on the 3D-TIs. It is generally noticed that asymmetric surface band structure in 3D-TIs between the electron and the hole sides frequently gives an asymmetric ambipolar action. These results confirm that high quality of massless Dirac cone surface states in BSTS thin films are well preserved even after the transfer process. When the film thickness was reduced to the 3 nm, $R_{xx}$-$V_G$ curve showed an ultrahigh on/off ratio ~ $10^4$ as shown in **Fig. 4(c)**. In the 3D-TIs, hybridization between top and bottom surface opened the energy gap on the topological surface states below the critical film thickness [35]. Since the $R_{xx}$ at $V_G = 0$ V is comparable to those in 3D-TI ultrathin film with film thickness of 2 - 3 nm[46], this high on/off ratio indicates the massive Dirac fermion in the present BSTS ultrathin film under the highly insulating electronic background.

As shown in **Fig. 4 (d)**, magnetoresistance for 18 nm thick film showed a weak anti-localization-like behavior decorated by clear Shubnikov–de Haas (SdH) oscillations at $V_G = 0$ V, manifesting that the transferred BSTS thin film is high quality. The oscillatory components after subtracting the background magnetoresistance as shown in Fig. **4 (e,f)** revealed clear $1/B$ periodic oscillations with their amplitude decay as a function of both $B$ and temperature $T$. The theoretical formula of SdH oscillations can be written as

$$\Delta R_{xx} = A exp(-\pi/\mu B) \cos[2\pi(B_F/B + 1/2 + \beta)] \quad (1),$$

where A is the amplitude, μ is the carrier mobility, $B_F$ is the periodic frequency of SdH oscillations, and $\beta$ is the Berry phase shift [47-49]. Analyses of the oscillatory components can allow us to confirm the phase shift in SdH oscillations caused by the finite Berry's phase, which is essential for the evidence of the nontrivial topological surface. Based on the fast Fourier transformation as shown in the **Fig. 4(g)**, two oscillatory components of $B_F = 8$ T and 21 T were extracted. The two Landau level (LL) fan diagram plots for $B_F = 8$ T and 21 T shown in **Fig. 4(h)** gave the $\beta$ values of 0.5 and 0.41 as the intercept. Although the value in 21 T oscillations were a little bit smaller than the ideal value of 1/2 theoretically expected in the massless Dirac fermions, similar deviation is frequently observed and explained in terms of the influence of the symmetrically curved Dirac dispersion [25, 40, 48, 49]. These two nontrivial topological states can reasonably be ascribed to both top and bottom surface Dirac states. In fact, the 21 T oscillation was almost erased whereas the 8T oscilation was remained under $V_G$ = -120 V, being consistent with the nonequivalent Fermi level shift by the electric field effect (**See supporting information Section 5 and Fig. S8, S9**). The carrier mobilities of the both surface Dirac fermions were evaluated by analyzing the SdH oscillations given by the equation (1) to be $\mu(B_F = 8$ T$) = 2500$ cm$^2$V$^{-1}$S$^{-1}$ and $\mu(B_F = 21$ T$) = 5100$ cm$^2$V$^{-1}$S$^{-1}$. These high carrier mobilities are comparable with those in the high quality bulk BSTS single crystal [12]. In order to estimate the cyclotron mass ($m^*$), temperature dependence of the amplitude of SdH oscillations was analyzed by the following equation

$$\Delta R_{xx} \propto \frac{\alpha T/\Delta E_N(B)}{\sinh[\alpha T/\Delta E_N(B)]} \quad (2),$$

where $\Delta E_N = heB/2\pi m^*c$ and $\alpha = 2\pi^2 k_B$ with the Landau index of $N$. For evaluation, we chose the most prominent intense signals with LL indices of $N_{21T}$ = 3, 3.5, 4 and $B_F$ = 21T and $N_{8T}$ = 2 for $B_F$ = 8T, and their fitting results are shown in **Fig. 4(i)**. These analyses yielded $m^*$ = (0.129±0.01)$m_e$ for $N_{21T}$ = 3.5, $m^*$=(0.125±0.03)$m_e$ for $N_{21T}$ = 3 and $m^*$ = (0.117±0.005)$m_e$, where $m_e$ is the free electron mass, and these values are almost same as those in bulk BSTS single crystals [49].

In summary, the epitaxial growth of BSTS on a mica substrate by the catalyst-free physical vapor deposition followed by the nondestructive film transfer process can accurately manipulate centimeter-size ultrathin films of high quality 3D-TI BSTS. Ambipolar electronic nature of Dirac fermions were confirmed for the transferred thin film by FETs fabricated on a $SiO_2$/Si substrate, where one observed clear SdH oscillations with π Berry phase as the consequence of the well-preserved topological surface. Carrier mobility estimated to be about 2500 – 5100 $cm^2V^{-1}S^{-1}$ is comparable to those of high quality bulk BSTS single crystal [12]. For the ultrathin region (Film thickness ~ 3 nm), an ultrahigh on/off ratio ~ $10^4$ was detected in $R_{xx}$-$V_G$ curve. This ultrahigh on/off ratio is consistent with the emergence of the massive Dirac fermion in the ultrathin regime of the present BSTS film and is merit for the application in the ultralow dissipative switching device[50]. Therefore, the present method involving the ultrathin film growth and the film transfer process will give a superior methodology for utilizing the novel surface electrical transport of 3D-TIs for various electric device applications in the future.

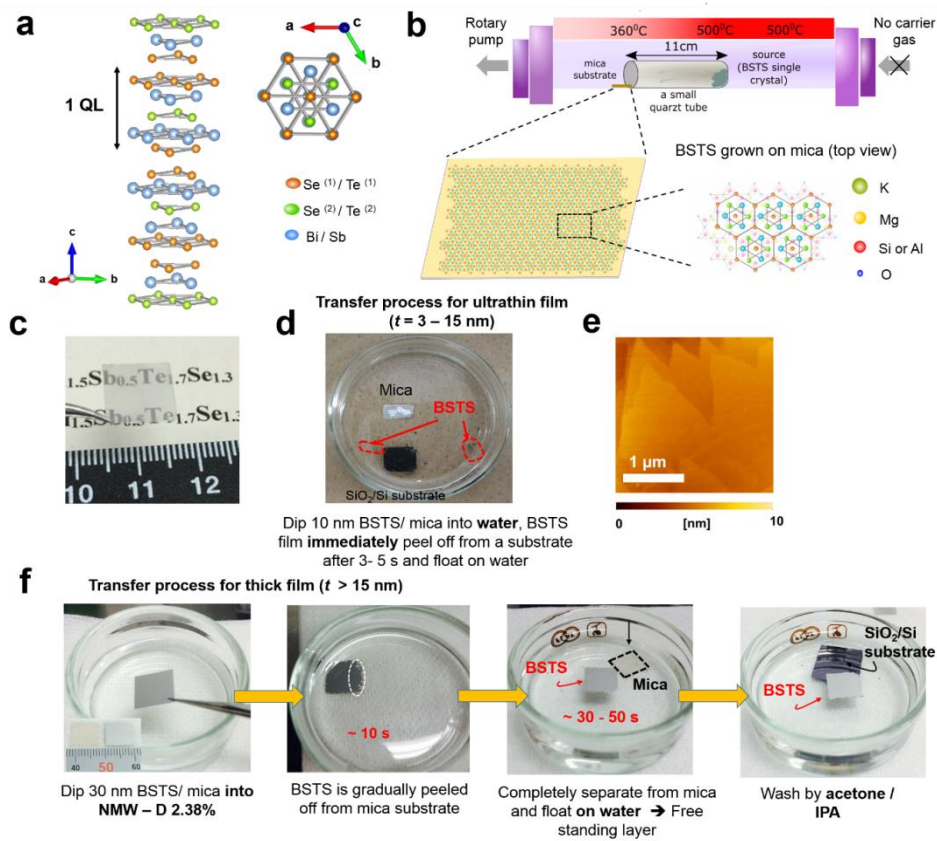

**Figure 1.** Catalyst - free physical vapor deposition (CF - PVD) of ultra-thin $Bi_{1.5}Sb_{0.5}Te_{1.7}Se_{1.3}$ (BSTS) films on mica and their transfer processes. (a) The crystal structure of BSTS. (b) Schematic view illustrating the present thin film crystal growth by employing two tubes with different diameters, and the proposed epitaxial growth mode on mica. (c) Photograph of 8 nm BSTS film with full coverage on mica. Photographs of a transfer process of BSTS film grown on mica to a $SiO_2$/Si substrate for (d) $t = 3 - 15$ nm and (e) AFM image of a 8 nm film transferred to a $SiO_2$/Si substrate, exhibiting atomically flat surface and large triangular terraces with no any damages. (f) $t > 15$ nm. BSTS films were dipped into deionized water in several to several tens of seconds then the completely peeled BSTS thin film floats on a water surface as a free–standing film. For $t > 15$ nm, as-grown BSTS thin films were precedently dipped into a 2.38 % tetra-methyl ammonium hydroxide (NMD–W 2.38%) aqueous solution in several seconds

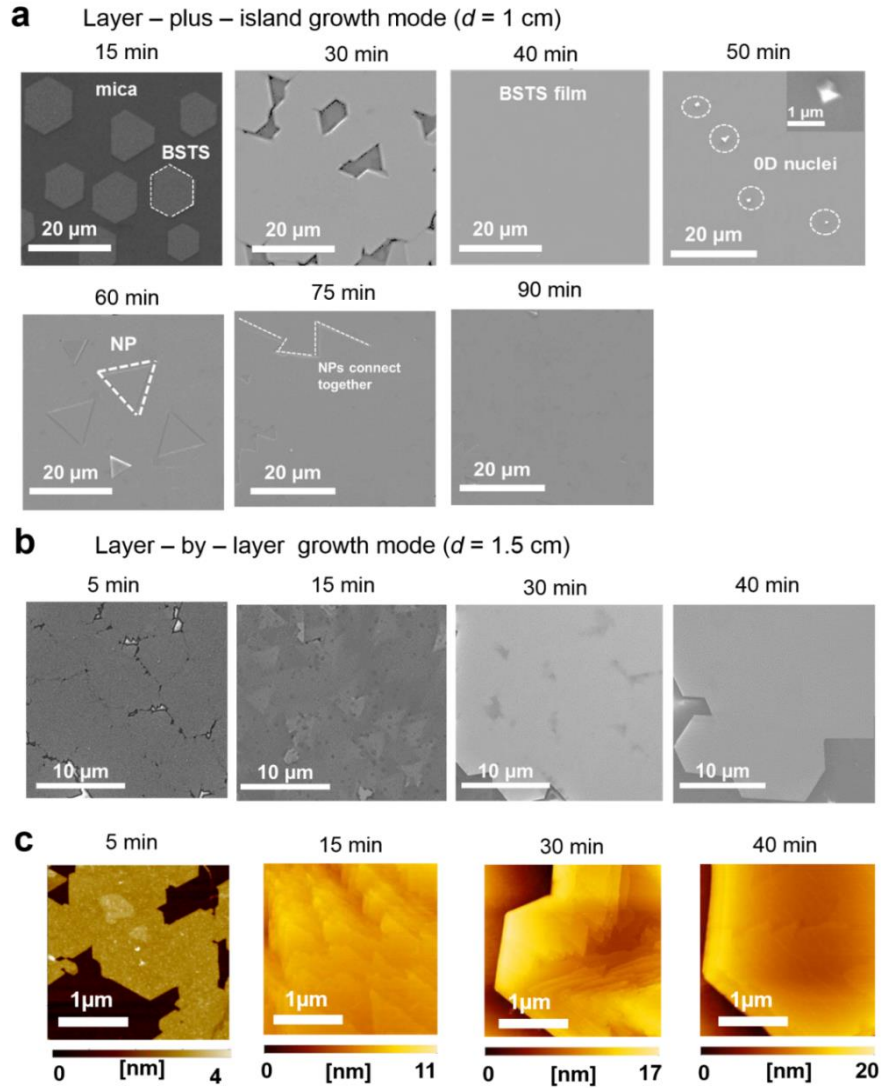

**Figure 2.** Thin film growth of large-area and uniform BSTS films in two different mechanisms. (a) Scanning electron microscopy (SEM) images of the layer–plus–island a growth mode ($d$ = 1cm). After 15 min growth, hexagonal nuclei (~ 10μm) with the same orientation were found. By increasing the growth time (~ 30 min), the small NPs enlarged in the lateral direction and connected together without grain boundaries (~40 min). The second layer was formed as the same mechanism when the growth time was increased (~ 50 – 90 mins). (b) SEM images in the layer– by–layer growth mode ($d$ = 1.5 cm). A 3 nm layer was obtained after 5 min. This layer is composed of many small NPs of various two- dimensional NPs (~ 1 – 2 μm in the lateral size). Depending on the growth time, various thin films of 8 nm (10 min), 12 nm (15 min), 18 nm (30 min) were grown. (c) AFM images of as-grown BSTS ultrathin films obtained in a different time after growth ($d$ = 1.5 cm), which illustrate the layer-by-layer growth mode. (20μm × 20μm views were shown in the Supporting Information Fig. S3 and S4)

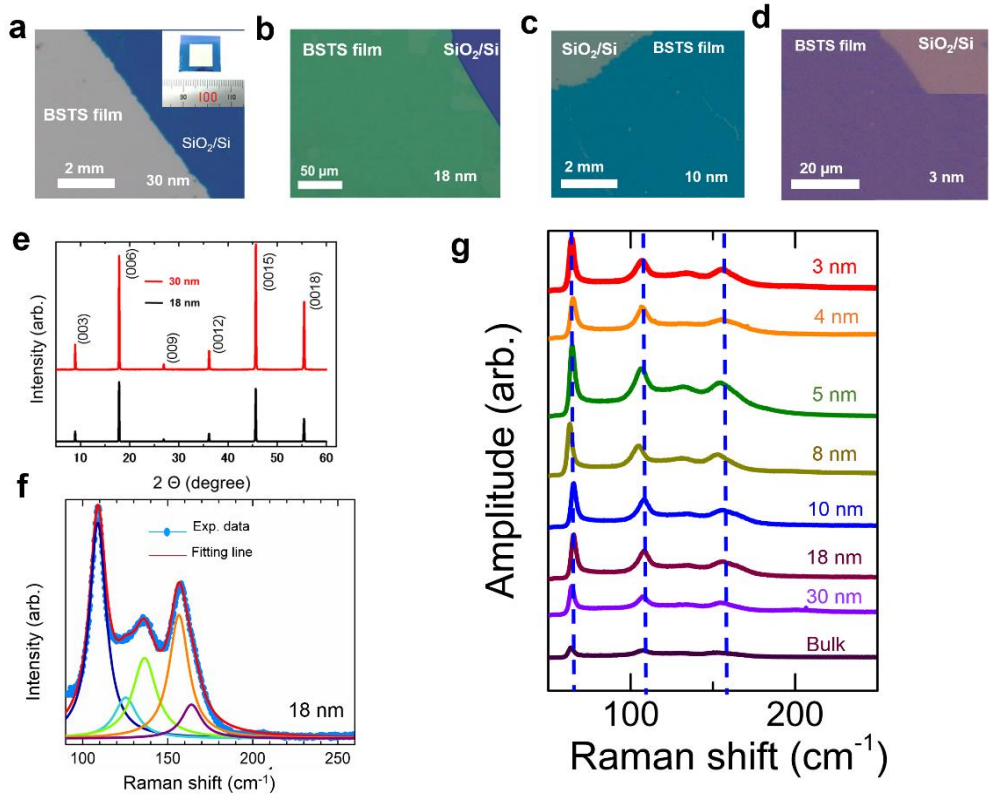

**Figure 3.** Characterization of the transferred films. (a) – (d) Optical image of 3 - 30 nm BSTS films after transferred to SiO$_2$/Si substrates. The excellent films with uniform color contrast indicate the transferred films are uniform. (e) X-ray diffraction of the transferred films on glass substrates with 18 and 30 nm in thickness. Raman spectra of transferred thin film. (f) Deconvoluted fittings using Lorentzian functions for a 18 nm thick film. The fitting displays five peaks at 109, 125, 136, 156 and 163 cm$^{-1}$. (g) Thickness dependence of Raman spectra of the transferred thin films and bulk BSTS single crystal [40] as the reference. The peak positions did not change among the different thicknesses from 3 nm to 30 nm.

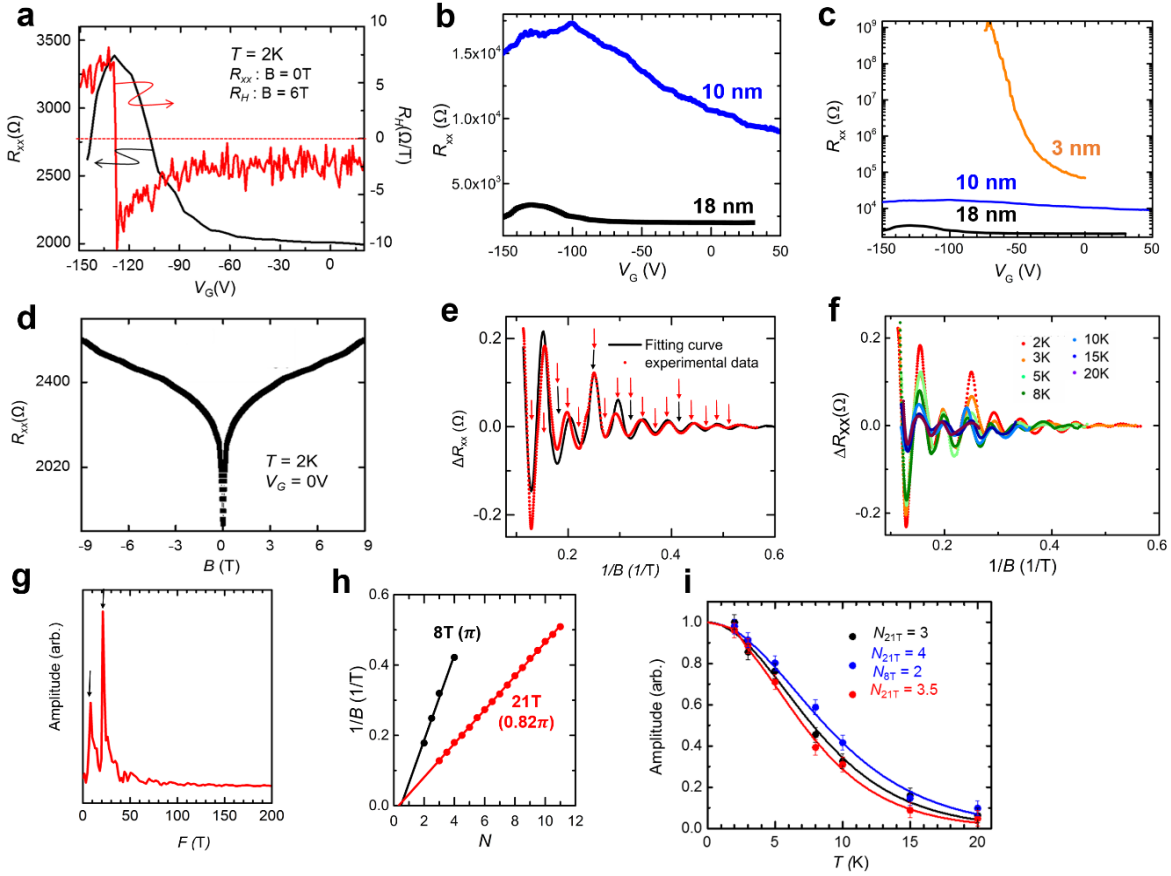

**Figure 4.** Electric transport properties of the BSTS film transferred on a SiO$_2$/Si substrate at 2 K. (a) Ambipolar field effect conduction of 4-probe resistance $R_{xx}$ and Hall coefficient $R_H$ for the 18 nm BSTS film as a function of $V_G$. At $V_G$= -130V, $R_{xx}$ approaches to a peak (~3400Ω) where the total carrier density becomes minimum. $R_H$ changes its sign at the Dirac neutral point when the dominant carrier type changes to the other type. (b, c) $R_{xx}$ - $V_G$ curves for the 3 - 18 nm BSTS films. (d) Magnetoresistance for $V_G$ = 0 V at 2 K. Under low magnetic fields, a weak antilocalization (WAL) like magnetoresistance was observed, whereas clear oscillations can be confirmed under high magnetic field. (e, f) Dependence of oscillation in $R_{xx}$ ($\Delta R_{xx}$) on 1/$B$ after subtracting the background. Solid lines are theoretical fittings by referring to the results of fast Fourier transformation as shown in (g). (h) Landau level (LL) fan diagram plot of $\Delta R_{xx}$. Solid lines are the linear fitting: $n = (F/B) + \alpha$ with $\alpha$ = 0.41 for $B_T$ = 21T and $\alpha$ = 0.50 for $B_T$=8T. The inset shows the fast Fourier transformation of $\Delta R_{xx}$ as a function of 1/$B$ at 2 K. (i) Temperature dependence of amplitude of the SdH oscillations for the LL indices of $N$ = 3(21 T), $N$ = 4 (21 T), 2 (8 T), $N$ = 3.5 (21 T). Solid lines are theoretical fittings.

AUTHOR INFORMATION

**Corresponding Author**

*Email: ytanabe@m.tohoku.ac.jp

**Email: tanigaki@m.tohoku.ac.jp

**Author Contributions**

Y.T. designed the project. N.H.T., Y.T., Y.S. S.Y.M. L.H.P. prepared $Bi_{2-x}Sb_xTe_{3-y}Se_y$ (BSTS) thin film by the catalyst free physical vapor deposition. N.H.T., Y.T. K.K.H. performed characterization of BSTS thin film. N.H.T, Y.T. and K.T. wrote the manuscript. All authors discussed the results and commented on the manuscript.

**Notes**

The authors declare no competing financial interest.

ACKNOWLEDGMENT

We are grateful to H. Shimotani and T. Kanagasekaran for instructions of a fabrication of a bottom-gate film transistor. We also grateful S. Ikeda, R. Kumashiro and K. Saito (Common Equipment Unit of Advanced Institute for Materials Research, Tohoku University) to support various experiments. This work was sponsored by the JSPS KAKENHI Grant Number JP15K05148.

ASSOCIATED CONTENT

**Supporting Information.**

Additional description and figure for materials and methods, influence of an inner tube on the growth, XPS studies on BSTS thin film, AFM images for BSTS thin film, EDX and Raman spectra for BSTS thin film, thickness variation of the electronic states of BSTS ultrathin film, electric field effect on the quantum oscillations were supplied as Supporting Information.